\documentclass[10pt,letterpaper]{article} 

\usepackage{opex3}
\usepackage{amsmath}
\usepackage{graphicx}
\usepackage{psfrag}
\usepackage[nosub]{psfragx}
\usepackage{cite}

\newcommand{\CU}{\ensuremath{^{1}}}
\newcommand{\NIST}{\ensuremath{^2}}
\newcommand{\CASA}{\ensuremath{^3}}
\newcommand{\PSU}{\ensuremath{^4}}
\newcommand{\HabWorlds}{\ensuremath{^5}}
\newcommand{\NISTGB}{\ensuremath{^6}}
\newcommand{\scottemail}{\ensuremath{^7}}

\begin{document}

\title{Demonstration of on-sky calibration of astronomical spectra using a 25 GHz near-IR laser frequency comb}
\author{Gabriel~G.~Ycas,\CU$^{,}$\NIST$^{,*}$\,
Franklyn~Quinlan,\NIST\, 
Scott~A.~Diddams,\NIST$^{,}$\scottemail\,
Steve~Osterman,\CASA\,
Suvrath~Mahadevan,\PSU$^{,}$\HabWorlds\,
Stephen~Redman,\NISTGB\,
Ryan~Terrien,\PSU\,
Lawrence~Ramsey,\PSU$^{,}$\HabWorlds\,
Chad~F.~Bender,\PSU$^{,}$\HabWorlds\, 
Brandon~Botzer,\PSU and\,
Steinn~Sigurdsson\PSU
}

\address{
\CU{}Department of Physics, University of Colorado, Boulder, CO, USA\\
\NIST{} National Institute of Standards and Technology, 325 Broadway, Boulder, CO, USA\\
\CASA{} University of Colorado, Center for Astrophysics and Space Astronomy, Boulder, CO, USA\\
\PSU Department of Astronomy and Astrophysics, Pennsylvania State University, University Park, PA, USA\\
\HabWorlds Center for Exoplanets and Habitable Worlds, Pennsylvania State University, University Park, PA, USA\\
\NISTGB{} National Institute of Standards and Technology, 100 Bureau Drive, Gaithersburg, MD, USA\\ 
\scottemail sdiddams@boulder.nist.gov\\
%$^*$\href{mailto:ycasg@colorado.edu}{ycasg@colorado.edu}\\
$^*$\textcolor{blue}{\underline{ycasg@colorado.edu}}\\

}

\begin{abstract}
We describe and characterize a 25 GHz laser frequency comb based on a cavity-filtered erbium fiber mode-locked laser. The comb provides a uniform array of optical frequencies spanning 1450 nm to 1700 nm, and is stabilized by use of a global positioning system referenced atomic clock. This comb was deployed at the 9.2 m Hobby-Eberly telescope at the McDonald Observatory where it was used as a radial velocity calibration source for the fiber-fed Pathfinder near-infrared spectrograph. Stellar targets were observed in three echelle orders over four nights, and radial velocity precision of $\sim$10 m/s ($\sim$6 MHz) was achieved from the comb-calibrated spectra.
\end{abstract}


\bibliographystyle{osajnl}

\noindent
\section{Introduction}

Searching for extrasolar planets with the Doppler radial velocity (RV) technique \cite{2010exop.book...27L} has been very successful. By measuring the tiny Doppler shifts in the light from stars induced by the presence of orbiting planets, more than 600 planets have been discovered\footnote{\texttt{http://www.exoplanet.eu}}. While this success illustrates the power of radial velocity surveys as a tool for planet-finding, it remains a significant technical challenge to reduce systematic uncertainties to the point of detecting an Earth-analog planet. Continued advances in the precision determination of stellar radial velocities via spectroscopic techniques will require significant observation time with a large-aperture telescope, an optimized fiber-fed high-resolution spectrograph, and stable calibration sources. Here we describe first experiments in the near infrared (IR) bringing together these main components, with a focus on the technical details, advantages, and challenges associated with the use of the laser frequency comb calibrator.

It is now recognized that laser frequency combs (LFC) offer new opportunities for the highest precision astronomical spectroscopy measurements (e.g. 1 cm/s RV, $3\times 10^{-11}$ fractional, or $\sim$6 kHz at 1550 nm) by providing a broad bandwidth, precisely tunable calibration spectrum that has an absolute accuracy traceable to the SI second\cite{2007MNRAS.380..839M,Li2008,OstermanSPIE2007,2008EPJD...48...57B,Steinmetz2008,springerlink:10.1007/s10686-011-9232-7}. The uniformly spaced, bright, and narrow features of the LFC spectrum are ideal for wavelength calibrations, while the absolute traceability of the comb allows for comparison of observations made on timescales from days to years and even from different observatories. Additionally, the tunability of the comb facilitates the probing of the fine details of a spectrograph's exact spectral response, as demonstrated in a recent characterization of the HARPS spectrograph's focal-plane array\cite{2010MNRAS.405L..16W}.

While several LFC-calibrators have been demonstrated in the visible spectrum\cite{Benedick2010,Stark2011}, there are compelling reasons to observe in the near IR. Because of their relatively low temperature, M dwarf stars, especially M4 and later, are faint in the visible but bright in the near IR (900 nm -- 1800 nm.) These stars are very good candidates for a radial velocity survey; M dwarf stars make up more than 60 \% of the stars within 10 pc of the Earth\footnote{RECONS census \texttt{http://www.recons.org}}, and due to their lower mass and luminosity, the radial velocity signal induced by an Earth-like planet orbiting the liquid water habitable zone of such a star is more than an order of magnitude larger than the signal induced by the Earth orbiting the Sun. 

These scientific motivations, taken together with the availability of established erbium fiber laser technology, make the H-band near IR atmospheric window (1.5 $\mu$m to 1.8 $\mu$m) a particularly attractive region for a LFC calibrator. In this spectral region there are no well-developed calibration sources similar to the thorium-argon lamps and iodine cells employed in the visible. While progress has been made with discharge lamps\cite{0067-0049-195-2-24} and gas cells \cite{0004-637X-692-2-1590}, these calibrators still exhibit the known challenges of irregular line spacing and line strength, barren regions, thermal background, line blending, and undefined aging. A LFC can overcome these issues and ultimately enable precision near IR calibrations and required instrument characterization at or below the 1 m/s level.

%{Should this paragraph be eliminated?\\
%% Unlike the majority of the planets detected, such potentially habitable planets have low masses, comparable to that of the Earth, and orbit at relatively large distances from their host stars. To maximize the size of the radial-velocity signal from a potentially habitable planet, there is interest in observing stars less massive than the Sun, such as M dwarfs. Due to their lower mass and luminosity, the radial velocity signal induced by an Earth-like planet orbiting the habitable zone of such a star is more than an order of magnitude larger than the signal induced by the Earth orbiting the Sun. Additionally, M dwarf stars are plentiful in our neighborhood of the galaxy, making up more than 60\% of the stars within 10 pc of the Earth\footnote{RECONS census \texttt{http://www.recons.org}}.
%}

In this paper, we provide the experimental details of recent trial measurements in which a 25 GHz-spaced laser frequency comb was used as the wavelength calibrator for astronomical spectroscopy in the range of 1.55 $\mu$m to  1.65 $\mu$m. Significantly, these results represent the first combination of all major components required for LFC-calibrated precision spectroscopy in the near IR, along with the first extraction of stellar RVs from this approach. Several stars were observed over a few nights with the 9.2 m Hobby-Eberly telescope, and the starlight and LFC were simultaneously coupled to the Pathfinder spectrograph\cite{Ramsey2010,2011AAS...21740101M}. Radial velocities were extracted from a limited number of spectral features with residual precision of $\sim$10 m/s, on the first test with a room-temperature near IR spectrograph, highlighting the promise of these calibrators.

%Suvrath et al think this reaches a bit too far
%a value which is competitive with the best existing near IR RV measurements\cite{Figueira2010}.

\section{Generation of 25 GHz laser frequency comb\label{section:lfcgeneration}}

The calibration laser frequency comb (LFC) is generated from the filtered spectrum of a 250 MHz passively mode-locked erbium fiber laser\cite{MenloRef}, similar to the one built and characterized in Ref. \cite{quinlan:063105}. The frequency of each optical mode is determined by the comb equation,
\begin{equation}
  \label{eq:comb}
  f_n = n\times f_\textrm{rep} + f_\textrm{ceo},
\end{equation}
where $f_\textrm{rep}$ is the laser's repetition rate, $f_\textrm{ceo}$ is the carrier-envelope offset frequency, and $n$ is an integer on the order of $10^{6}$. To frequency stabilize the mode-locked laser (MLL), the frequencies $f_\textrm{rep}$ and $f_\textrm{ceo}$ must be detected and locked. The most straightforward detection of $f_\textrm{ceo}$ requires an octave-spanning spectrum\cite{DavidJ.Jones04282000}. Using an Er:fiber amplifier and polarization-maintaining highly nonlinear fiber\cite{HiranoHNLF}, $f_\textrm{ceo}$ is detected and locked to 160 MHz by modulation of the MLL's pump laser diode. The repetition rate, $f_\textrm{rep}$, is directly detected and locked by controlling the laser cavity length. By referencing the synthesizers to which $f_\textrm{ceo}$ and $f_\textrm{rep}$ are locked to a global positioning system-disciplined rubidium clock, the frequencies of the laser modes are fixed with a fractional uncertainty of $10^{-10}$, limited by the accuracy of the clock.

\begin{figure}[htpb]
  \begin{center}
    \includegraphics[width=\textwidth]{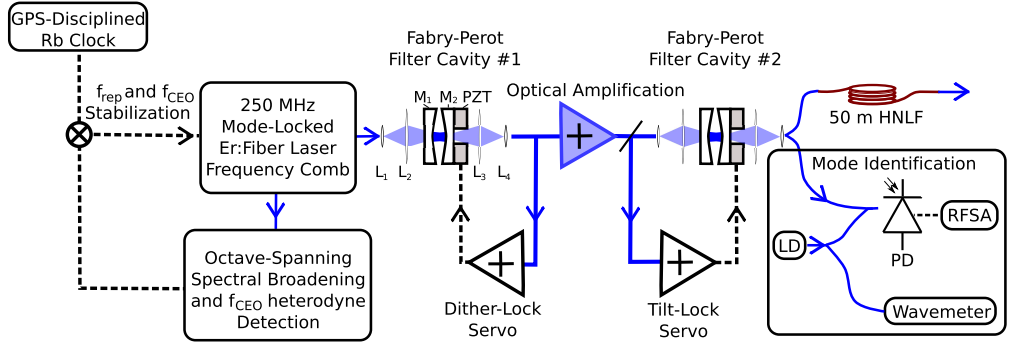}
  \end{center}
  \caption{Schematic of the laser frequency comb. The 250 MHz passively mode-locked erbium fiber laser is stabilized by locking the repetition rate $f_\textrm{rep}$ and carrier-envelope offset frequency $f_\textrm{ceo}$ to a global-positioning system (GPS) disciplined rubidium clock. Light from the mode-locked laser is sent through the Fabry-Perot cavity (mirrors $M_1$ and $M_2$) with mode-matching between the cavity and single-mode fiber provided by lenses $L_1-L_4$. The transmitted light is amplified to 1.4 W, then sent through a second, identical Fabry-Perot cavity. The pulse is re-compressed and spectral broadening is achieved using 50 m of highly-nonlinear fiber (HNLF). Mode identification is achieved by measuring the beat of a single comb mode with a wavemeter-calibrated CW laser diode (LD) using a photodiode (PD) and radio-frequency spectrum analyzer (RFSA).}
  \label{fig:laserdrawing}
\end{figure}

For the calibration of the  resolution $\lambda/\Delta\lambda=50,000$ Pathfinder astronomical spectrograph we chose to use a comb mode-spacing of 25 GHz, yielding one comb line every 6.5 resolution elements. To generate a spectrum with this spacing using our 250 MHz mode-locked laser, the Fabry-Perot mode-filtering scheme\cite{1989IJQE...25...97S,2009OptL...34..872K,2008OptL...33..959C} is employed. Two identical filter cavities are constructed, using mirrors with reflectivity of 99.8\,\% and radii of curvature 5 cm and 10 cm spaced by $\sim{}0.6$ cm for a free-spectral range of 25 GHz and finesse of approximately 2000. To provide fine, high-speed length tuning of the cavity length, one mirror in each cavity is glued to a ring-shaped piezoelectric transducer (PZT). The cavities are coupled to standard single-mode fiber (SMF) using pigtailed collimators and two additional lenses for mode-matching, allowing  a fiber-to-fiber coupling efficiency of $\sim{}20\,\%$.

\begin{figure}[htbp]
\begin{centering}
\includegraphicx[width=0.9\textwidth](%x
\psfrag{s03}[bc][bl][1][180]{\color[rgb]{0,0,0}\setlength{\tabcolsep}{0em}\renewcommand{\arraystretch}{1}\hspace{10ex}\begin{tabular}{c}\vspace{4ex}Power per Mode\end{tabular}}%
\psfrag{s04}[bc][bl]{\color[rgb]{0,0,0}\setlength{\tabcolsep}{0pt}\renewcommand{\arraystretch}{1}\hspace{-10ex}\begin{tabular}{c}Power per Mode (dBm)\end{tabular}})%
{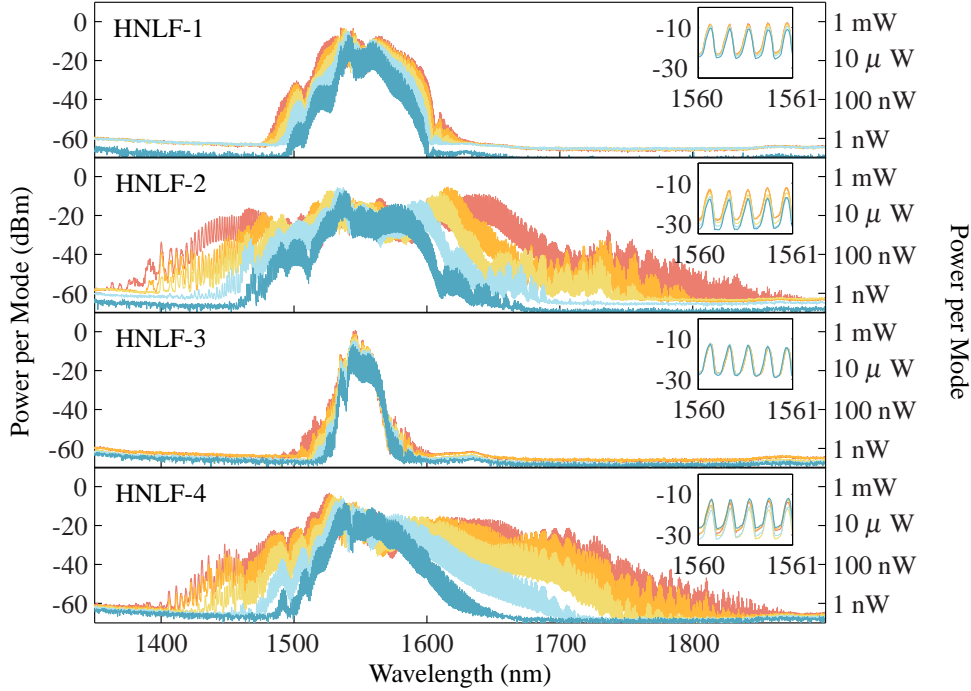}
\end{centering}
\caption{\label{fig:hnlfs}Supercontinuum spectra generated with 25 GHz pulses in a variety of highly nonlinear optical fibers (HNLFs) at different launch powers. Insets: zoom-in between 1560 nm and 1561 nm, showing the resolved 25 GHz-spaced comb modes. The dynamic range of the measurement, and any apparent asymmetry in the lines, is limited by the optical spectrum analyzer used for the measurement. At the input of the fiber, the pulses have a duration of 300 fs, as determined by nonlinear autocorrelation. The energy per pulse is varied from 3 pJ (narrow spectra, foreground) to 8 pJ (broad spectra, background), corresponding to 80 mW--200 mW average power. The lengths and dispersion parameters of the fibers are HNLF-1: 100 m, -0.14 ps/nm/km at 1550 nm, HNLF-2: 50 m, +0.3 ps/nm/km at 1550 nm, HNLF-3: 48 m, +6.7 ps/nm/km at 1550 nm, HNLF-4: 100 m, +2.5 ps/nm/km at 1550 nm.}
\end{figure}

Generation of the 25 GHz calibration spectrum proceeds as in Ref. \cite{quinlan:063105} and is illustrated in Fig. \ref{fig:laserdrawing}. The two identical filter cavities are used to select a subset of modes with spacing of 25 GHz, and cascaded optical amplifiers placed between the cavities provide the gain required to attain the 300 mW of power required for nonlinear broadening of the LFC. After filtering and amplification, the 25 GHz comb light is compressed to a $\sim$300 fs pulse using dispersion-compensating fiber and launched into highly nonlinear fiber (HNLF)\cite{Okuno1999}. A number of different HNLFs were tested, and spectra recorded at different input powers are reported in Fig. \ref{fig:hnlfs}. For our input pulse, the best results were achieved by use of 50 m of the ``HNLF-2'' fiber for generation of the calibration spectrum. Finally, to resolve the ambiguity in the mode-number of the filtered comb modes a heterodyne measurement between a single mode of the filtered comb and a wavemeter-calibrated continuous-wave (CW) laser is performed. 

\section{Optical heterodyne measurement of 25 GHz comb\label{section:opticalheterodyne}}
%% Good words
After nonlinear broadening, the comb spans from 1450 nm to 1700 nm, providing a calibration spectrum with extremely high signal-to-noise ratio suitable for an astronomical spectrograph covering the H-band. The frequency of each mode of the filtered 25 GHz comb is known with an accuracy of $10^{-10}$, limited by the GPS-disciplined rubidium standard. While this corresponds to $\sim$3 cm/s uncertainty, the actual achievable precision is in most cases limited by other sources. Here we discuss the most significant sources of uncertainty related to our method of comb generation. In particular, we extend the characterization of the LFC line-shape, absolute frequency accuracy, and side-mode suppression from ref. \cite{quinlan:063105} with optical heterodyne measurements to assess the uncertainties that arise from spurious and unsuppressed side-modes and technical ambiguities in servo control.

One of the chief sources of uncertainty is a result of the use of nonlinear spectral broadening in conjunction with a Fabry-Perot cavity filtered laser frequency comb. Each filter cavity suppresses the two nearest-neighbor 250 MHz comb modes on either side of the 25 GHz comb modes by $10^3$ per cavity, and while the use of two cavities increases the suppression to $10^6$, four-wave mixing processes taking place in the optical amplifiers and nonlinear fiber transfer power back into the suppressed modes. Depending upon the pulse and fiber parameters during the nonlinear broadening process, the parametric gain has the potential to be so large for the suppressed modes to have amplitude comparable to and even larger than the nominal 25 GHz comb mode\cite{quinlan:063105,Chang:10}. It is also possible that, in the cavity filtering and broadening processes, a spurious 250 MHz comb mode, for example a mode lying on a higher-order transverse mode of the filter cavity\cite{Steinmetz2009,quinlan:063105,Siegman}, could see amplification and lead to a skewing of the apparent line center.

\begin{figure}
  \begin{center}
    \includegraphicx[width=\textwidth]{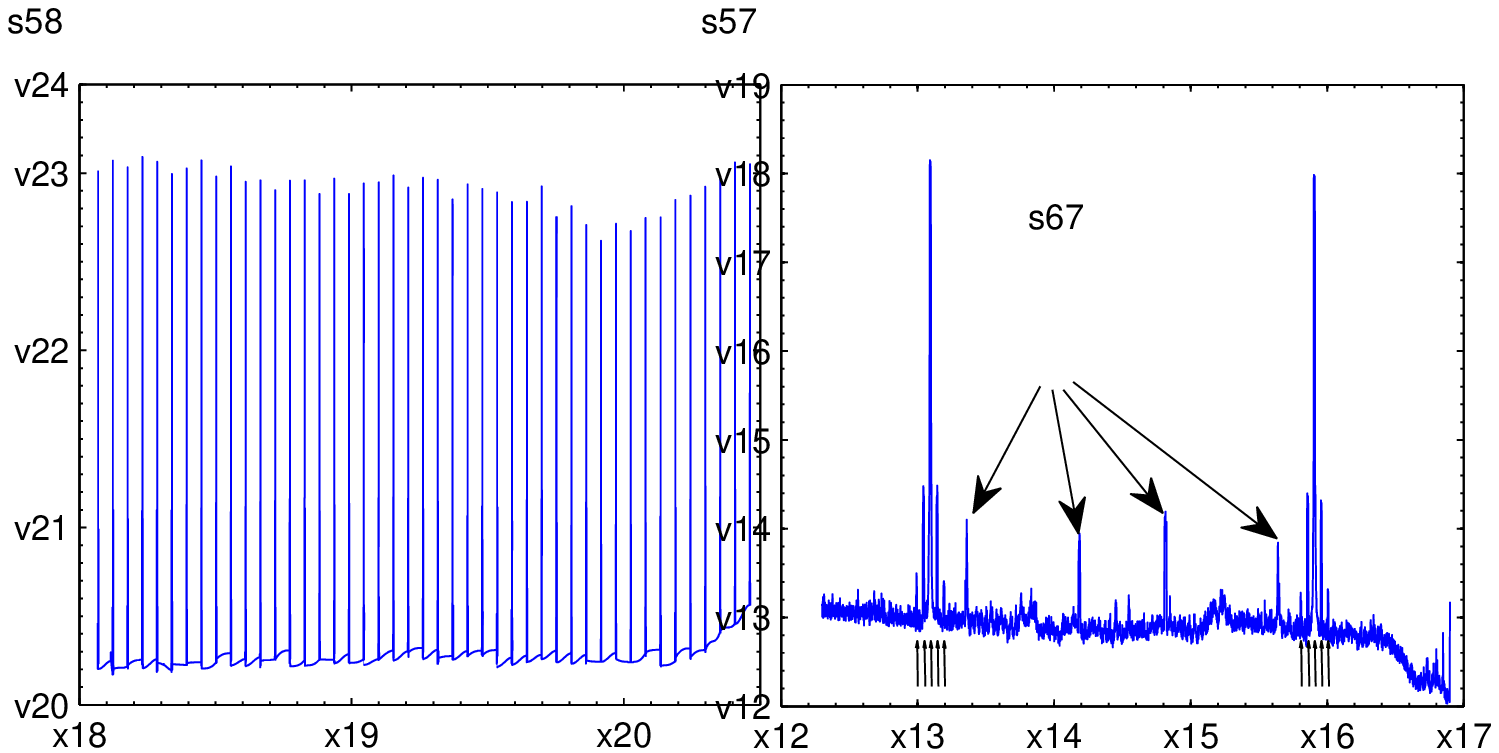}
    \vspace{0.5cm}
    \caption{\label{fig:rf_optical_spect} a.) Optical spectrum obtained by analyzing 45 individual heterodyne measurements of the calibration spectrum with a tunable CW laser. Within the 50 dB - 55 dB dynamic range of this measurement, no spurious optical modes between comb teeth were detected other than nearest-neighbor modes offset by 250 MHz and 500 MHz. The noise floor indicated is an estimate, defined by the the mode of the detected RF power from each measurement from 1-25 GHz. b.) Radio-frequency spectrum from heterodyne of LFC with CW laser tuned near 1621 nm. Peaks determined to be in the optical spectrum are marked from below by arrows.
}
  \end{center} 
\end{figure}

To verify that the comb's optical spectrum is free of spurious modes, a heterodyne measurement continuously covering 1610 nm to 1620 nm was made by tuning a CW laser across the 25 GHz LFC. Because the amplitudes of the side-modes are determined by the action of nonlinear processes, and not by the transfer function of the filter cavities, a measurement of the filter cavity dispersion \cite{Li:10} is not sufficient and heterodyne measurements after nonlinear broadening are required. The results of this measurement, derived from 90 individual radio-frequency spectra, are shown in Fig. \ref{fig:rf_optical_spect}. This measurement demonstrates that no spurious peaks exist between the 25 GHz comb modes above the -55 dB noise floor between 1610 nm and 1620 nm. Because the spurious effects from higher-order transverse modes are expected to occur across the comb and spurious modes due to four-wave mixing processes are expected to most strongly present far from the center of the filtered LFC, this measurement assures us that the spurious modes observed by the Pathfinder spectrograph are negligible or well-understood.

\begin{figure}[htb]
\begin{center}
\includegraphicx[width=0.9\textwidth](%
\psfrag{s02}[bc][bl]{\color[rgb]{0,0,0}\setlength{\tabcolsep}{0pt}\renewcommand{\arraystretch}{2.6}\begin{tabular}{c}Fractional Shift of Apparent Center\end{tabular}}%
\psfrag{s07}[bc][bl][1][180]{\color[rgb]{0,0,0}\setlength{\tabcolsep}{0pt}\hspace{-6ex}\begin{tabular}{c}Side-mode Suppression (dB)\end{tabular}}%
){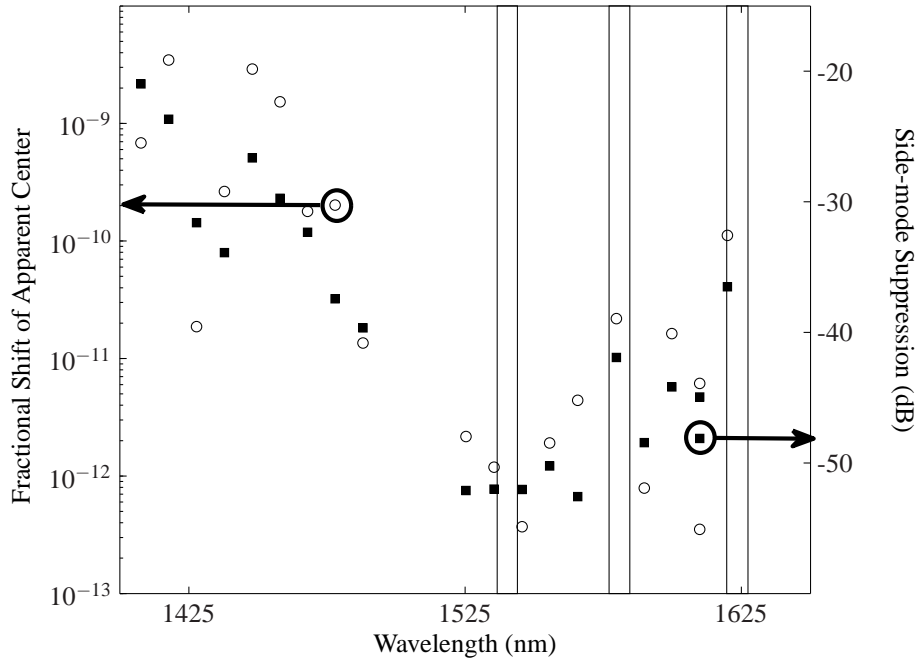}
\end{center}
\caption{\label{fig:smsr10nm}Measured apparent shifts (circles) of 25 GHz comb mode centers and side-mode suppression (squares), derived from optical heterodyne measurements. Data were taken by measuring the heterodyne beat of a tunable CW laser with single modes of broadened comb from 1400 nm to 1625 nm and apparent shifts were calculated by using the weighted average of the comb mode and nearest-neighbor side-modes. Enclosed in rectangles are the three wavelength regions observed by the Pathfinder spectrograph. Data were taken with CW laser tuned to both the high- and low-frequency sides of each mode, and the side mode amplitudes from the two measurements were averaged. Note that the line-center shift, which depends upon not only the degree of side-mode suppression but also on asymmetry, mirrors the side-mode suppression.}
\end{figure}

While there are no large spurious modes in the 25 GHz spectrum, incompletely suppressed side modes are present in the spectrum\cite{2008EPJD...48...57B,Steinmetz2008,Chang:10,Li:10,quinlan:063105}. The Pathfinder spectrograph's point-spread function has a full-width at half maximum of $\sim 3$ GHz, and as a result the apparent line-center is the weighted average of the main comb mode and the nearest-neighbor side modes,

%\begin{equation}
%  \label{eq:smshift}
%  f_\textrm{apparent} = \frac{f_0 + A_- (f_0 - 250\, \textrm{MHz})+ A_+ (f_0 + 250\, \textrm{MHz})}{1+A_-+A_+},
%\end{equation}
\begin{equation}
  \label{eq:smshift}
  f_\textrm{apparent} = \frac{\sum_{i=-N}^{N} A_i(f_0+i \times 250\, \textrm{MHz})}{\sum {A_i}}
\end{equation}

where $A_i$ are the relative powers of the comb modes, $f_0$ is the optical frequency of the central mode, and N is 50; the fractional shift of the line center is $(f_0-f_\textrm{apparent})/f_0$. For side-mode levels measured, only the central mode and two nearest-neighbor $\pm 250$ MHz modes significantly affect the apparent line center.  To measure the size of this shift in the calibration spectrum, heterodyne measurements with a CW laser were made at 10 nm intervals, from 1400 nm to 1620 nm, recording the relative side-mode amplitudes at each point. The results of this measurement, shown in Fig. \ref{fig:smsr10nm}, demonstrate that near 1540 nm the side-mode suppression is highest at $>50$ dB and that moving to both higher and lower wavelengths sees a significant decrease in suppression. Nonetheless, in the spectral bands of interest, the apparent spectral shift is $\leq 10^{-10}$ fractionally.

\begin{figure}
  \begin{center}
    \includegraphicx[width=0.8\textwidth](
\psfrag{s05}[bc][bl]{\color[rgb]{0,0,0}\setlength{\tabcolsep}{0em}\renewcommand{\arraystretch}{1}\hspace{0ex}\begin{tabular}{c}\small Optical Power\\Transmitted by\\Cavity (AU)\\\vspace{-1em}\end{tabular}}%
\psfrag{s09}[bc][bl]{\color[rgb]{0,0,0}\setlength{\tabcolsep}{0em}\renewcommand{\arraystretch}{1}\hspace{0ex}\begin{tabular}{c}\small Side-Mode\\Suppression (dB)\\\vspace{-1em}\end{tabular}}%
\psfrag{s13}[bc][bl]{\color[rgb]{0,0,0}\setlength{\tabcolsep}{0em}\renewcommand{\arraystretch}{1}\hspace{0ex}\begin{tabular}{c}\small Side-Mode\\ Asymmetry (dB)\\\vspace{-1em}\end{tabular}}%
\psfrag{s17}[bc][bl]{\color[rgb]{0,0,0}\setlength{\tabcolsep}{0em}\renewcommand{\arraystretch}{1}\vspace{-1cm}\begin{tabular}{c}\small Fractional Shift of\\Apparent Center\\\vspace{-1em}\end{tabular}}%
){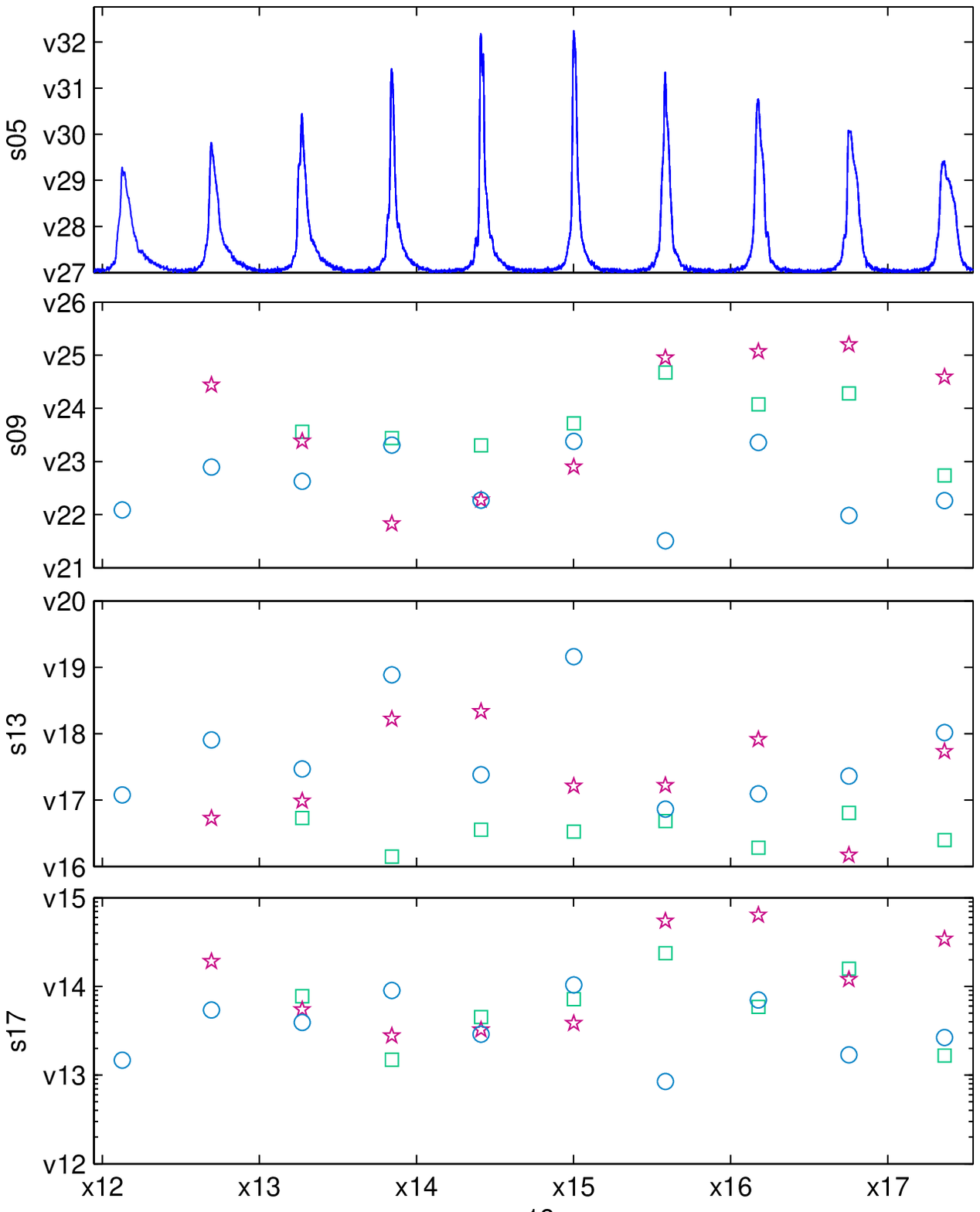}
  \end{center}
  \caption{\label{fig:varlockpt}Measured effect of the lock point of the first filter cavity on the supercontinuum after the HNLF. The first filter cavity was locked to 10 different transmission peaks (top) and the side-modes were measured at 1439 nm (stars), 1566 nm (circles), and 1625 nm (squares.)  The measurement shows that the choice of lock point has no strong influence on the side-mode suppression, asymmetry, or shift of line center. The uncertainty in these measurements is $\pm$ 2 dB, due to amplitude noise of the frequency comb and frequency noise of the CW laser.}
\end{figure}

One final source of uncertainty arises from the technical issue of locking the filter cavities to transmit a particular set of 250 MHz modes. Because of cavity dispersion, only a few such sets are efficiently transmitted, but there is freedom to choose among these sets of modes. By measuring the side-mode suppression of each set of transmitted modes, see Fig. \ref{fig:varlockpt}, it is confirmed that equal side-mode suppression, and thus accuracy, is obtained regardless of which set of transmitted modes is chosen.

\section{Calibration of the Pathfinder spectrograph\label{section:HET}}

After testing in the NIST laboratories in Boulder, Colorado, the laser frequency comb-based calibrator was transported to the McDonald Observatory in southwest Texas where it was used to calibrate the Pathfinder spectrograph at the Hobby-Eberly telescope (HET). The Pathfinder spectrograph is a prototype fiber-fed near-IR spectrograph with resolution $\lambda/\Delta \lambda=50,000$ operating from 1-1.8 $\mu$m\cite{2008PASP..120..887R,Ramsey2010}. The spectrograph is assembled on an optical breadboard inside a passively stable room-temperature enclosure with a liquid nitrogen cooled HgCdTe detector array having 1024 $\times$ 1024 pixels on an 18.5 $\mu$m pitch (HAWAII-1 HgCdTe Astronomical Wide Area Infrared Imager.) To reduce the thermal background from the warm optics, cascaded cold edge-pass filters inside the detector dewar are used to reject out-of-band radiation. Three 300 $\mu$m core multimode fibers, potted in epoxy and polished one atop the other, feed the spectrograph. One fiber carries light from the telescope, while the other two are used for calibration. The spectrograph's resolution is set by use of a 100 $\mu$m entrance slit, cross-dispersed and imaged into 4.4 pixels of the focal-plane array. For the demonstration of the LFC calibrator, the angles of the spectrograph gratings were adjusted away from their designed Y-band positions for operation in the H-band. In this configuration, fractions of three echelle orders were imaged onto the focal-plane array, covering a total of 22.5 nm of spectral bandwidth between 1537 nm and 1627 nm. The drift of the spectrograph was measured during the run using both uranium-neon lamp lines and the LFC, and is typically hundreds of meters per second per day. Because the telescope and calibration fibers closely track each other, this drift has been shown to limit RV precision only at levels below 3 m/s \cite{2008PASP..120..887R}.

\begin{figure}[htbp]
  \begin{centering}
  \includegraphicx[width=\textwidth]{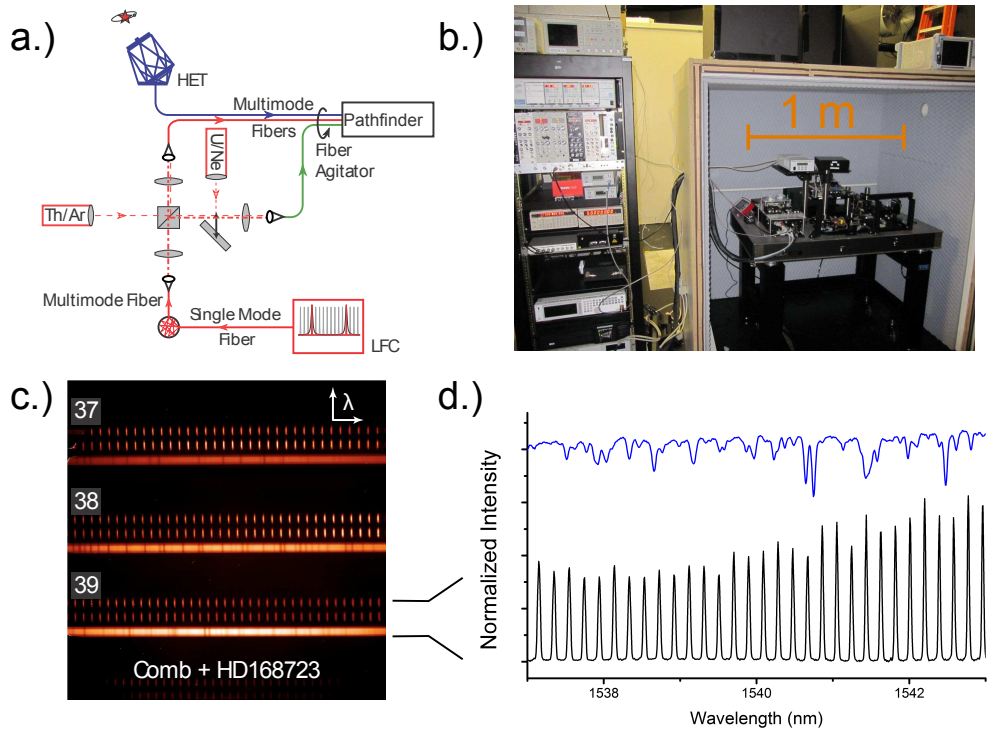}
   \end{centering}
  \caption{\label{fig:hetfig} a.) Schematic of the coupling of light sources into Pathfinder spectrograph. Free-space beam splitters direct the different calibration sources -- LFC, U/Ne, and Th/Ar -- into the two calibration fibers. A commercial paint shaker is used as a fiber agitator to mitigate modal noise from the multimode fibers. 
b.) A photograph of the laser frequency comb calibrator in the spectrograph room at the Hobby-Eberly telescope at the McDonald Observatory. The electronics rack, left, houses the laser drivers, electronics for servo loops, and other electronics. The optics breadboard, right, holds the mode-locked laser, $f-2f$ interferometer, two filter cavities, and fiber-optic components. The entire optics breadboard is enclosed within a wood-foam acoustic enclosure.
c.)  Image from focal-plane array of Pathfinder spectrograph showing light from both the laser frequency comb and the star HD168723. Echelle orders 37, 38, and 39 are visible, with wavelengths of 1536.6 nm to 1543.9 nm, 1577.1 nm to 1584.6 nm , and 1619.7 nm to 1627.4 nm. d.)  Line out showing the 25 GHz laser frequency comb over a single echelle order.
}
\end{figure}

The laser frequency comb, which consists of a 2.5 ft$\times$4 ft optical breadboard floating on vibration-isolating air legs and 19 inch electronics rack (see Fig. \ref{fig:hetfig}(b)), was set up in the spectrograph room of the HET, adjacent to the thermal and acoustic enclosure housing Pathfinder. Calibration light was coupled from the single-mode fiber output of the LFC into a 300 $\mu$m core fiber using a 4 inch diameter PTFE (Teflon) integrating sphere with an internal baffle. Although the power loss through the integrating sphere was roughly a factor of $10^6$, there was sufficient light available for the spectrograph. The 300 $\mu$m fiber output from the integrating sphere was then sent to the Pathfinder calibration bench, where an arrangement of beamsplitters (see Fig. \ref{fig:hetfig}(a)) allowed for the illumination of the spectrograph's two 300 $\mu$m calibration fibers with combinations of the LFC and hollow-cathode U/Ne and Th/Ar lamps. Using this capability, an \textit{in situ} measurement of the spectrum from 1454 nm to 1638 nm of a hollow cathode U/Ne lamp was made using the LFC and Pathfinder spectrograph\cite{2011arXiv1112.1062R}. The data for this measurement were collected in only a few hours, demonstrating the strength of the LFC as a calibration tool for astronomical spectroscopy.

An additional configuration allowed for the coupling of comb light into a multimode fiber routed to a position near the prime focus of the HET, where it could illuminate a screen used for flat-field calibration of the science fiber. The high power of the LFC combined with the various configurations of illumination of the three spectrograph fibers provided a means of cross-checking the quality of spectra obtained using the different spectrograph fibers. While not studied in detail in these preliminary experiments, we envision taking advantage of this flexibility in the future to examine and reduce systematic effects related to time-varying pointing errors, non-uniform mode excitation (modal noise), and the speckle-type pattern from the coherent LFC calibrator. 

In the present experiments it was seen that modal noise \cite{2001PASP..113..851B,grupp2003}, which is the changing illumination of the detector resulting from the interference of the finite number of excited optical modes in the 300 $\mu$m fibers, limited the attainable RV precision to $\sim$ 10 m/s. To mitigate the effect of modal noise, we employed the integrating sphere for coupling the single-mode laser light into the 300 $\mu$m optical fiber and a commercial paint mixer for active fiber agitation. In spite of these efforts, modal noise still dominated all other limitations on RV precision, such as point-spread function changes due to the pupil illumination of the telescope, detector response, the signal-to-noise ratio, and the number of stellar lines. More details about the impact of modal noise on measurements with the LFC are presented in section 5.1 of Ref. \cite{2011arXiv1112.1062R}. The particular issue of coupling single spatial mode frequency comb light to the multimode fibers and spectrograph is a new aspect of this problem which will require further investigation.

\begin{figure}[!h]
  \begin{center}
    \includegraphicx[width=\textwidth]{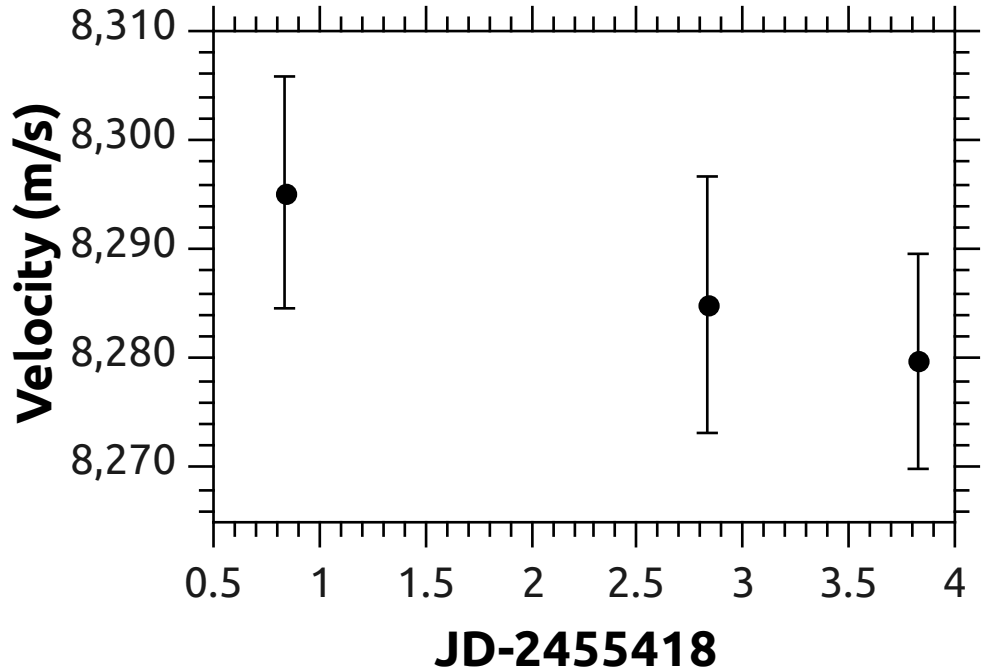} 
  \end{center}  
  \caption{\label{fig:rvplots}Radial Velocities were calculated from one order for the star Eta Cas using a binary mask cross-correlation similar to that described in Ref. \cite{1996A&AS..119..373B}, with the correlation mask generated from an NSO FTS atlas of the solar spectrum. We only included the deep stellar absorption lines, well separated from telluric features, as part of the correlation. This returns not only the relative velocity of Eta Cas for every night, but also the `absolute velocity' of the star since the wavelength solution is referenced to the frequency comb, and the mask is referenced to the solar velocity. The `absolute velocity' we measure is entirely consistent with the velocities for Eta Cas reported by Nidever et al. \cite{0067-0049-141-2-503} (8314 m/s), exhibiting only a $\sim$25 m/s difference even though  different analysis techniques and entirely different stellar lines at very different wavelengths are used in both cases.  The error bars shown in the figure are photon noise limited error bars, calculated using the theoretical prescription of Bouchy, Pepe, and Queloz\cite{2001A&A...374..733B}.}
\end{figure}

The frequency comb was operated together with the Pathfinder instrument at the HET over a two week period. Over this time, the calibration spectrum had a constant envelope and power output, and the filter cavities could be locked to transmit the same 25 GHz subset of modes. The long-term stability of the comb freqencies was determined by the Rb clock, which was measured at NIST both before and after the experiment to have fractional absolute accuracy of $1\times10^{-10}$.

During the observation run, 648 spectra were recorded with 5 minute integration times, of which 91 were of astronomical objects, including the stellar targets HD16873, Sigma Draconis, Vega, Eta Cassiopeae, and Upsilon Andromeda. Useful spectra were obtained on three nights with good weather and seeing, and three nights with variable conditions. During these measurements, the comb operated without fault, although adjustments to the servos were required before each night. A typical spectral image for HD16873 is shown in Fig. \ref{fig:hetfig}(c) and extracted spectra are presented in \ref{fig:hetfig}(d). As seen in the figures, the comb provides a uniform grid of calibration markers with high signal-to-noise for each echelle order. The simultaneously recorded science spectra show both absorption features intrinsic to the stellar atmosphere as well as sharp telluric CO2 lines from the earth's atmosphere.  Radial velocities of the known stable star Eta Cassiopeae, obtained with the comb as the simultaneous reference to track instrument drift, are shown in Fig. \ref{fig:rvplots}.  The derived uncertainty of each 5 minute exposure was $\sim$ 30 m/s, and typically 6-11 such exposures were acquired back-to-back. The scatter in the data about the mean for the back-to-back exposures was $\sim$ 28 m/s, consistent with the uncertainty attributed to each exposure. The dominant limitation on precision is again modal noise, and is was seen to average down when the back-to-back exposures were combined. This indicates that the modal noise is not static, but changes on a time-scale similar to the 5 minute exposure time. 

All of the data in Fig. \ref{fig:rvplots} are derived from the analysis of a single echelle order, and in spite of modal noise limitations show an H-band RV precision of $\sim$ 10 m/s, a level competitive in the NIR (with an uncooled instrument testbed). While improvements in data reduction are ongoing and full details of the RV analysis will be provided in a future publication, these preliminary results are very encouraging and motivate the further development and optimization of high-precision infrared frequency comb spectroscopy.

\section{Conclusion}
In summary, we present the first reported calibration of stellar spectra using a laser frequency comb and a high-resolution fiber-fed infrared astronomical spectrograph. In laboratory measurements, the comb was shown to have absolute accuracy of better than $2\times10^{-10}$ from 1450 nm to 1630 nm, corresponding to a radial velocity precision of better than 6 cm/s. Using the laser frequency comb as calibrator, observations of stellar targets were made with the Pathfinder astronomical spectrograph, allowing for radial velocity precision of 5-15 m/s. The lessons learned in this work provide the basis for development of a LFC for the new, facility-class Habitable Zone Planet Finder\cite{HZPF2010} spectrograph, covering the  Y, J, and H bands from 0.95 $\mu$m to 1.70 $\mu$m. The combined advances in laser technology and astronomical spectrographs should enable 1 m/s calibration precision in the infrared, which will provide a new tool for exoplanet surveys of the numerous population of M dwarf stars.

\subsection*{Acknowledgments}

We acknowledge support from NIST, from NASA through the NAI and Origins grant NNX09AB34G, and from the NSF grants AST-0906034, AST-1006676, AST-0907732, and AST-1126413. This work was partially supported by funding from the Center for Exoplanets and Habitable Worlds, which in turn is supported by the Pennsylvania State University, the Eberly College of Science, and the Pennsylvania Space Grant Consortium. FJQ acknowledges the support of the National Research Council. The Hobby-Eberly Telescope (HET) is a joint project of the University of Texas at Austin, the Pennsylvania State University, Stanford University, Ludwig-Maximilians-Universit\"{a}t M\"{u}nchen, and Georg-August- Universit\"{a}t G\"{o}ttingen. The HET is named in honor of its principal benefactors, William P. Hobby and Robert E. Eberly. We thank the HET resident astronomers (John Caldwell, Steve Odewahn, Sergey Rostopchin, and Matthew Shetrone) and telescope operators (Frank Deglman, Vicki Riley, Eusebio ``Chevo'' Terrazas, and Amy Westfall) for their expertise and support. We also thank the HET engineers and staff who provided us with critical  assistance during the day: Edmundo Balderrama, Randy Bryant, George Damm, James Fowler, Herman Kriel, Leo Lavender, Jerry Martin, Debbie Murphy, Robert Poenisch, Logan Schoolcraft, and Michael Ward. This work would not be possible without them. The authors thank M. Hirano of Sumitomo Electric Industries for providing HNLFs 1, 2, and 3.

\end{document}